\documentclass[aps,pra,superscriptaddress,twocolumn]{revtex4}
\usepackage{amsmath}
\usepackage{calligra}
\usepackage{amsfonts}
\usepackage{graphicx}
\usepackage{dcolumn}
\usepackage{stackrel,amssymb}
\usepackage{latexsym}
\usepackage{color}
\usepackage{ulem}
\usepackage[squaren,thickspace,thickqspace]{SIunits}
\usepackage{bbding}
\usepackage{bm}

\begin{document}

\title{Multi-ion sensing of dipolar noise sources in ion traps}

\author{F. Galve}
\email{fernando@ifisc.uib-csic.es}
\affiliation{IFISC (UIB-CSIC), Instituto de F\'isica Interdisciplinar y Sistemas Complejos, Palma de Mallorca, Spain}
\author{J. Alonso}
\email{alonso@phys.ethz.ch}
\affiliation{Institute for Quantum Electronics, ETH Z\"urich, Otto-Stern-Weg 1, 8093 Z\"urich, Switzerland}
\author{R. Zambrini}
\affiliation{IFISC (UIB-CSIC), Instituto de F\'isica Interdisciplinar y Sistemas Complejos, Palma de Mallorca, Spain}

\begin{abstract}
Trapped-ion quantum platforms are subject to `anomalous' heating due to interactions with electric-field noise sources of nature not yet completely known. 
There is ample experimental evidence that this noise originates at the surfaces of the trap electrodes, and models 
assuming fluctuating point-like dipoles are consistent with observations, but the exact microscopic mechanisms behind anomalous 
heating remain undetermined. Here we show how a two-ion probe displays a transition in its dissipation properties, enabling experimental access
to the mean orientation of the dipoles and the spatial extent of dipole-dipole correlations. This information can be used to test the validity of candidate microscopic models, which predict correlation 
lengths spanning several orders of magnitude. Furthermore, we propose an experiment to measure these effects with currently-available traps and techniques.

\end{abstract}

\maketitle

Trapped atomic ions constitute prominent candidates for deployable technologies 
exploiting the unintuitive properties of quantum mechanics 
\cite{09Home,17Linke}. A number of scalable architectures have 
been proposed \cite{02Kielpinski,00Cirac,17Lekitsch}, but technical constraints 
limit the current computational power of high-fidelity trapped-ion quantum 
machines to less than ten qubits 
\cite{16Debnath}. One key aspect towards the most notorious scalable schemes is 
trap miniaturization. This eases scalability and allows for faster quantum 
operations on the computational space (internal electronic states)
\cite{03Leibfried,11Ospelkaus} as well as the quantum bus (ions' motion) 
\cite{12Bowler,12Walther}. However, trapped-ion experiments suffer from motional 
heating due to interactions with noise sources
of origin not yet completely known \cite{00Turchette,blatt,wineland}. The measured 
effects of this so-called `anomalous' heating scale strongly with the inverse of 
the ion-electrode distance, posing a major
obstacle to trap miniaturization.\\
\indent Systematic experimental studies suggest that the origin of anomalous heating is 
due to contaminants on the surfaces of trap electrodes \cite{blatt}. In 
\cite{100-fold} the NIST ion-storage group treated electrode surfaces with 
ion bombardment. The 100-fold reduction in the observed heating rates points at 
adsorbates as probable culprits for the noise. But recent studies show that this 
is not the whole picture \cite{wineland2017},
suggesting that only electrode surfaces subject to radio-frequency drives (as 
required for ion trapping) are accountable for the heating. This result has a 
profound impact on the search for possible 
microscopic models since, to our knowledge, all previous studies considered 
thermally-driven processes (see \cite{blatt} for a review of proposed microscopic
models). In particular, the diffusion of 
adsorbates \cite{diffusion1,diffusion2,diffusion3,diffusion4}
is consistent with the most advanced surface-science experiments realized to 
date on a trap setup \cite{Sadeghpour}, appearing to be a plausible mechanism 
for anomalous heating.\\
\indent Up until now studies of the noise origin have focused exclusively on its scaling 
for a {\it single trapped ion}. This is usually assumed as $S_E~\sim \omega^{-\alpha} d^{-\beta} T^\gamma$
with $d$ the ion's distance to the electrode, $\omega$ its motional frequency and $T$ the
trap-electrode's temperature. For different microscopic models a strong distance scaling $\beta>3$ is predicted
and is consistent with experiments \cite{blatt,wineland}, whereas frequency and temperature dependencies
vary for different models. Here we propose a new way of measuring noise which 
can give {\it finer details} on its microscopic origin.\\ 
We show that for a trap holding two or more ions a noise crossover effect must take 
place: there is a point where the heating rates of the center-of-mass (COM) and relative motions equate.
This previously unknown crossover occurs when ions are separated by 
comparable distances from each other ($l$) and from the trap electrodes ($d$, see fig.~\ref{fig1}).
We will show that its characteristics depend not only on the spatial extent of correlations 
of dipole-dipole fluctuations $\xi$ (or the size of patches), but also on the average 
orientation of the dipoles. This phenomenon is in stark contrast to that found 
for an homogeneous lattice environment \cite{FerSciReps}, where the properties of the 
propagator (anisotropy, resonant manifold, etc.) determine the crossover 
characteristics.\\
\indent We derive estimates of the correlation lengths $\xi$ of 
dipole-dipole fluctuations for different microscopic models in the literature 
and show that the one-ion noise-level departs from its typical $d^{-4}$ scaling 
\cite{blatt} when $d\sim\xi$. Likewise the noise crossover 
for two ions is shifted to higher ion-ion distances $l$ in this regime. Further, 
the absence/presence of noise crossover for different motional degrees of 
freedom uniquely determines the mean orientation of dipoles. As a side result, 
we show that the effect of dipole orientations can also be observed with a single 
ion. This might explain recent results at NIST, where noise levels were
measured to be highest for ion motion parallel to the surface projection of the 
sputter beam at a well-defined angle \cite{PrivateHite}. 
Finally, we propose a realistic experiment to measure and characterize the noise 
crossover. This can be carried out with current state-of-the-art Paul traps and 
techniques.  

\section{Electric field noise} Every dipole source on the surface of an electrode 
represents a noisy source of electric potential at the ion position: 
$\phi(\vec{r})=\vec{\mu}\cdot\vec{r}/|\vec{r}|^ 3$,
with $\vec{\mu}$ the dipole moment of the ion and $\vec{r}$ the position of the 
dipole relative to the ion. Heating rates for two ions along 
a given motional degree of freedom, say $x$,
depend on the correlators $\langle E_x(\vec{r}_i,\tau)E_x(\vec{r}_j,0)\rangle$, 
with $E_x(\vec{r}_i,\tau)$ 
the total electric field component along $x$ at time $\tau$ at the ion's position 
$\vec{r}_i$. We derive (appendix \ref{appendix:appA}) the master 
equation for two ions in Lindblad form, exhibiting the desired heating rates.

\begin{figure}[b!]
\includegraphics[width=0.9\columnwidth]{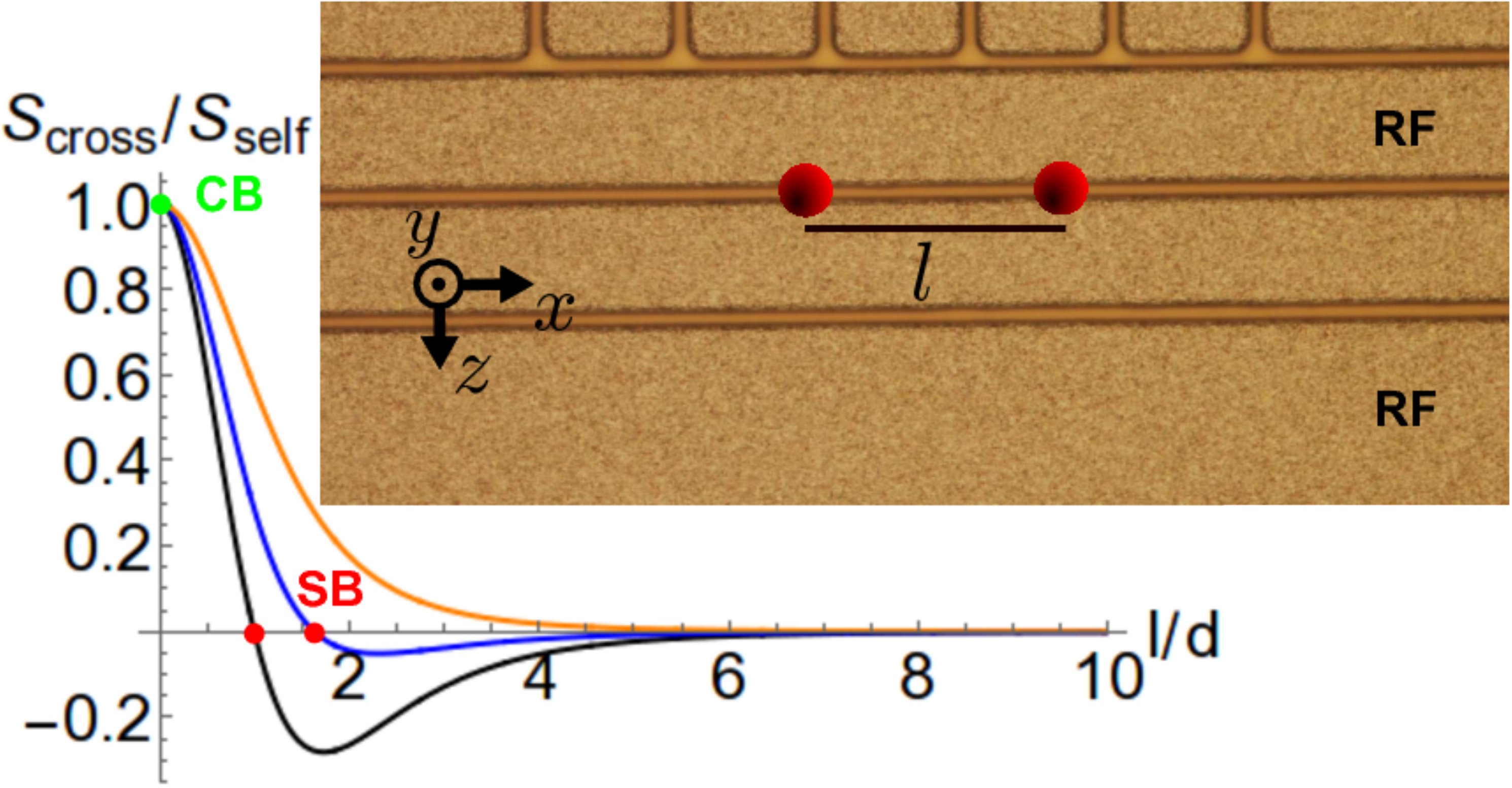}
\caption{Ratio $S_{\rm{cross}}/S_{\rm{self}}$ for the different motional degrees 
of freedom of a two-ion system: $x$ (black), $y$ (blue) and $z$ (orange). The dipoles are 
assumed to be uniformly distributed and pointing along $y$. The electrode-area included in
this simulation is a square of side $20d$, which is enough to avoid finite-size effects. 
Inset: Sketch of a surface-electrode Paul trap with segmented electrodes, similar to \cite{wineNAT2011}.
The ion-ion distance is $l$ and the ion-electrode distance is $d$, which is 
usually similar to the width of the central (horizontal) electrode.}
\label{fig1}
\end{figure}

Let us consider two coupled identical ions aligned along the $x$-axis (see Fig.~\ref{fig1}), 
separated by $l$ from each other and $d$ from the trap electrodes. We will 
assume in what follows that all relevant modes of motion have been cooled close 
to their respective ground states. It is convenient to move to a normal-mode 
picture $x_\pm=(x_1\pm x_2)/\sqrt{2}$, where $x_{1,2}$ are the ions' positions 
with respect to a lab reference frame. Here, the $(+)$ mode corresponds to the 
center-of-mass motion (COM) and the $(-)$ mode to relative motion (stretch 
mode), with eigenfrequencies $\Omega_\pm$. The heating rates of the normal modes are given by
\begin{equation}
\Gamma_\pm=\frac{e^2}{4m\hbar \Omega_\pm}S_\pm,
\end{equation}
with $e$ and $m$ the ions' charge and mass respectively, $\hbar$ the reduced Planck constant, $\Omega_j$ the normal-mode frequencies, and 
\begin{equation}
S_\pm=2\int_{-\infty}^\infty \text{d}\tau e^{-\text{i}\Omega_\pm 
\tau}\langle E_x^{(\pm)}(\tau)E_x^{(\pm)}(0)\rangle
\end{equation}
the electric-field fluctuations' spectral densities.
The fields acting on the normal modes are a linear combination of the fields 
seen by the individual ions: $\tilde{E}_x^{(\pm)}=[E_x(\vec{r}_1)\pm 
E_x(\vec{r}_2)]/\sqrt{2}$. Defining $s_{i,j}(\tau):=\langle E_x^{(i)}(\tau)E_x^{(j)}(0)\rangle$, we can 
write $S_{\pm}=\int_{-\infty}^\infty\text{d}\tau e^{-\text{i}\Omega_\pm 
\tau}[s_{1,1}+s_{2,2}\pm(s_{1,2}+s_{2,1})]$. 
Although environmental noise can lead to coupling between both normal modes if the ions are weakly coupled \cite{breuer}, 
a sufficiently homogeneous electrode guarantees that this coupling is negligible 
(see discussion in appendix \ref{appendix:appA}), leading to independent decay channels for the normal modes.
For the two ions, this translates into a {\it{self-noise}} for each ion ($S_{1,1}$, $S_{2,2}$) and a {\it{cross-noise}} 
($S_{1,2}$, $S_{2,1}$). The cross-noise governs the transition from common bath (CB) to separate baths 
(SB) \cite{breuer}, two emblematic dissipation scenarios in open quantum systems. The former dissipates only the coordinate $x_+$ and leaves $x_-$ 
unaffected, whereas the latter yields equal-rate dissipation for both. Regrouping $S_{\rm{self}}=(S_{1,1}+S_{2,2})/2$ and 
$S_{\rm{cross}}=(S_{1,2}+S_{2,1})/2$, normal modes $(\pm)$ dissipate as $S_{\pm}=S_{\rm self}\pm S_{\rm 
cross}$. SB occurs (no frozen mode) when $S_{\rm{cross}}=0$, and
CB (frozen $x_-$) when $S_{\rm{cross}}=S_{\rm{self}}$. We will 
later show that the counterintuitive `anti-common' bath case (aCB) where $S_{\rm{cross}}=-S_{\rm{self}}$ (frozen $x_+$) is also possible.

The $x$ component of the electric field at a position $\vec{r}$ is given by 
$E_x(\vec{r},t)=-\partial_x\phi(\vec{r})=\sum_i (1/4\pi\epsilon_0) 
\mu_i(t)g_x(\vec{r},\vec{r}_i)$, where $g_n(\vec{r},\vec{r}_i)$ are geometric 
functions (appendix \ref{appendix:appB}) which depend on the orientation of dipoles, and $\mu_i\hat{=}|\vec{\mu}_i|$. Thus, the cross- and self-noise are given by expressions
\begin{equation}
\label{eq3}
s_{i,j}(\tau)=\sum_{l,k}\frac{\left<\mu_l(t)\mu_k(0)\right>}{(4\pi\epsilon_0)^2} g_x(\vec{r}_i,\vec{r}_l)g_x(\vec{r}_j,\vec{r}_k),
\end{equation}
where $\left< \mu_l(t)\mu_k(0)\right>$ is a correlation function between dipoles 
$l$ and $k$. This dipole-dipole correlator 
features separated temporal and spatial terms for proposed microscopic models \cite{blatt}, so we 
can approximate it by $\left<\mu_l(t)\mu_k(0)\right>\simeq s_\mu(t)f(\vec{r}_l,\vec{r}_k)$ \cite{footnote}. Here, $f$ is a spatial correlation profile (it 
can be a phononic correlation decay in the electrode, 
a domain function for dipoles in the same patch, etc.) and the approximation is 
valid if time-fluctuations are similar across the whole surface. Thus, after 
Fourier integration ($\mathcal{F}$) we will have two main ingredients: the dipole fluctuation 
spectrum at the eigen-frequencies, $S_\mu(\Omega_\pm)\hat{=}\mathcal{F}[s_{\mu}](\Omega_\pm)$, and geometric 
contributions (from $f$ and $g$). For ions in separate wells the Coulomb 
coupling is small compared to the eigen-frequencies and
$S_\mu(\Omega_+)\simeq S_\mu(\Omega_-)$. The focus of our work will therefore be 
on the geometric part.

\section{Noise crossover} We start by considering the noise characteristics of an 
electrode containing dipoles which point normal to the surface. This is a 
typical assumption even in cases where microscopic details 
are calculated to a large extent \cite{Sadeghpour}. In order to give a clear 
picture of the origin of the crossover, let us consider the simplest case: all 
dipoles are pointing normal to the surface 
$\vec{\mu}_i=\mu_i \hat{u}_y\ \forall i$, they are uncorrelated, and we focus on 
the motional mode along $x$ as a function of $l$. In this case, the cross-noise 
is proportional to the sum $\sum_i g_x(\vec{r}_1,\vec{r}_i) g_x(\vec{r}_2,\vec{r}_i)$. In the simplest 
scenario we can assume the dipoles to be almost homogeneously distributed on the 
surface and replace the sum by an integral. Noting that (see e.g. \cite{haffner})
$$g_x(\vec{r}_1,\vec{r}_i)=\frac{d\, (x_i- x_1)}{((x_i-x_1)^2+z_i^2+d^2)^{5/2}},$$ 
one can see that the integral with respect to $z$ is always finite and positive. 
However, the integral with respect to the dipole coordinate $x$ has an M shape, meaning that for particular parameters the area enclosed by this shape 
will vanish (appendix \ref{appendix:appC}). The crossover originates 
precisely due to the fact that for a given combination $\{ d,l \}$ the 
cross-noise integral will be zero and the cross-noise changes sign. In 
figure~\ref{fig1} we see the ratio $S_{\rm{cross}}/S_{\rm{self}}$ for all three 
degrees of freedom. When $d\gg l$ we have $+1$ (CB), and around $d=l$ we have 0 
(SB). When $d<l$ we have negative values, which means that the stretch mode 
dissipates at a rate $S_-=S_{\rm{self}}-S_{\rm{cross}}$ higher than 
$S_+=S_{\rm{self}}+S_{\rm{cross}}$.
The crossover is absent for $z$-motion (in agreement with figure 27 of \cite{blatt}), however
for $x$- and $y$-motion it is present, leading to an aCB regime.

One could wonder whether a pure aCB regime is at all possible. We show in appendix \ref{appendix:appD}
that for a stylus-trap configuration \cite{09Maiwald} one can reach 
$S_{\rm{cross}}/S_{\rm{self}}$ ratios approaching -1. However, in such a trap at least one of the 
ions will be necessarily driven by micromotion \cite{98Wineland2}, which could 
make the aCB regime hard to observe.

\section{Dipole orientation} Before dealing with possible spatial correlations 
among dipoles, let us consider what happens when dipoles are not normal to the 
surface; to our knowledge this has never been considered before.
Since the effective dipole of the ion is given by 
its displacement from the radio-frequency (RF) null and along its motion, 
different dipole orientations can cause different noise levels along different 
directions. These can vary by a factor of 6 depending on dipole orientation (see appendix \ref{appendix:appE} and fig.~\ref{dipole1ion}), would be detectable with a one-ion probe, and might be 
behind recent observations with a single ion at NIST \cite{PrivateHite}. After 
treating the electrode surfaces with ion-beam sputtering, they observed that two 
orthogonal motional degrees of freedom were subject to different noise levels, suggesting a preferential orientation of surface dipoles which could be 
caused by the generation of Gold nanochannels on the treated surfaces. \\
In order to sense anomalous heating with 
two ions, it is convenient to have both ions at RF null zones. This can be 
achieved in segmented, linear Paul traps (fig. \ref{fig1}), where 
trap heights are typically $d\simeq L_z$, with $L_z$ the width of the 
central axial electrode. The ions can be placed at arbitrary positions along the 
extent of the linear section $L_x$, allowing for a tunable inter-ion separation 
$l$. In the coming analysis we will use as a reference the setup in 
\cite{wineNAT2011}. Note that this does not compromise the generality of our 
results. We will further assume that only RF electrodes are sources of noise,
corresponding to the upper and lower (long) electrodes in figure~\ref{fig1}.

\begin{figure}[h]
\includegraphics[width=0.85\columnwidth]{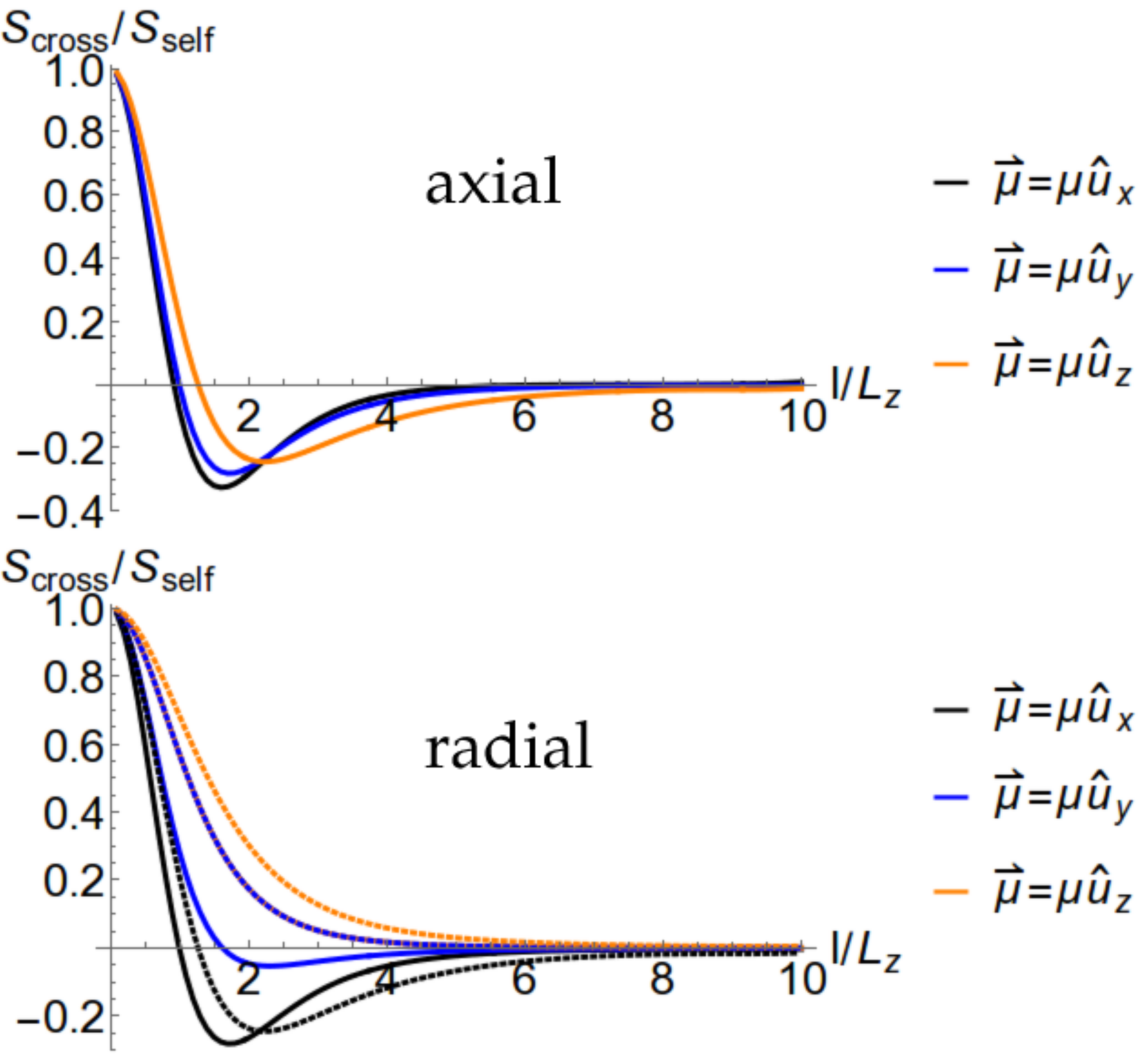}
\caption{Ratio of cross- to self-noise for uncorrelated dipoles homogeneously covering the surface of (RF driven) electrodes in the segmented planar trap of 
ref. \cite {wineNAT2011} (two RF electrodes of length $L_x$, one of them at positive $z$ with 
width $L_z\simeq L_x/10$, the other starting at negative $z=-L_z$ of width $2L_z$; see fig. \ref{fig1}). We set $d\simeq L_z$, and plot (top) the noise experienced by the axial motion ($x$), and 
(bottom) by radial motion along the $y$ (solid) and $z$ (dashed) axes. Note that the dashed blue line and the solid orange line overlap in this plot.}
\label{fig3}
\end{figure}

\begin{table}[b!]
\caption{\label{tab:table1} Truth table indicating the presence (\Checkmark) 
or absence (\XSolidBrush) of noise crossover for radial motion along the $y$ and $z$ axes for a given dipole orientation. 
This information together with the results of heating-rate measurements of the radial normal modes allow for discrimination of surface-dipole orientations.}
\begin{tabular}{|c|c|c|}
\hline
\textrm{Crossover {\it y}-motion}&
\textrm{Crossover {\it z}-motion}&
\textrm{Dipole orientation}\\ 
\colrule
\Checkmark & \Checkmark & $\mu_x$\\
\Checkmark & \XSolidBrush  & $\mu_y$\\
\XSolidBrush & \XSolidBrush  & $\mu_z$\\
\hline
\end{tabular}
\end{table}
Different dipole orientations result in different dependencies on $l$ of the self- and cross-noise terms (fig.~\ref{fig3}). 
Therefore, experimental measurements of the $S_{\rm{cross}}/S_{\rm{self}}$ ratio can reveal the mean orientation of the dipole
fluctuators. Note that the increased sensitivity of the radial modes as compared to the axial motion render the former as most 
suitable for this analysis. Table~\ref{tab:table1} can be used to gain qualitative insight about the mean orientation of the dipoles.
For arbitrary orientations the resulting curves lie between those plotted for the three principal axes, but the structure of crossovers 
in the table is still valid. For example, any orientation $\mu_r$ will keep the $\{$\Checkmark, \XSolidBrush$ \}$ signature for
any direction $r\neq z$ in the $yz$-plane, even if the crossover for $y$-motion is less steep \cite{note1}.

\section{Spatial dipole-dipole correlations}
Different microscopic models of dipolar fluctuations result in different two-point spatio-temporal correlation 
functions $\langle\mu_l(t)\mu_k(0)\rangle\simeq s_\mu(t)f(\vec{r}_l,\vec{r}_k)$, so it is natural to wonder whether
we can measure its spatial dependence $f$ with our scheme, and thus falsify given models.
For one trapped ion we find that the consequence of spatial correlations is the breakdown 
of the typical $d^{-4}$ scaling for the spectral noise density when 
$d\lesssim\xi$, tending towards $d^{-1}$. A simple mean-field argument (appendix \ref{appendix:appF}) explains the 
saturation and value of this scaling. For two trapped ions, spatial correlations translate
into a shift of the crossover point to higher $l/L_z$, except for the case of $y$-motion 
with $\mu_x$ pointing dipoles, where the crossover can even disappear. Importantly, 
however, these effects are also appreciable only for $\xi\sim L_z$ (appendix \ref{appendix:appG}).\\
\indent We derive next the size $\xi$ of correlations for several proposed models and discuss the possibility to probe them.
For patch models the dipoles are electronic cloud deformations at the 
surface due to different crystallographic orientations 
of domains in the electrode metal, so the function $f$ satisfies $f=1$ whenever two 
dipoles lie on the same domain and $0$ otherwise. Patch sizes in the range 
$[\unit{10}{nm},\unit{10}{\micro m}]$ have been measured \cite{patches}, so ions at 
distances of tens of microns could potentially feel effects of big enough 
patches.
Another proposed model is based on adatoms (or molecules) stuck to the electrode 
surface with their induced dipole fluctuating through phononic thermal noise 
\cite{safavi}. In such model the dipoles of two adatoms would be spatially
correlated through a phonon manifold resonant with their motional bound states 
($\approx$\unit{300}{GHz} for Neon on Gold) \cite{safavi}. Taking the dispersion 
relation of Gold ($\approx$\unit{5}{THz} at $\lambda\approx \unit{21}{pm}$, \cite{gold}), such
frequency would correspond to wavelengths $\sim \unit{1}{nm}$. This indicates that 
correlations would decay at distances $\xi$ on the order of 
nano-meters, possibly of $\sim\unit{100}{nm}$ for heavier adsorbed molecules, 
still far away from the scale of tens of microns. If we assume that dipole fluctuations are
RF-driven \cite{wineland2017}, then a drive at $\sim$\unit{100}{MHz} would correspond
to wavelengths of $\sim \unit{1}{\micro m}$ (provided a viable mechanism relating 
RF and phonon excitations exists). For the model of adatoms 
diffusing on the surface, the spatial scale for dipole-dipole correlations is 
$\sqrt{D/\Omega_\pm}$, with $D$ the diffusion constant. Considering a range 
$D\in\unit{[10^{-14},10^{-10}]}{m^2 Hz}$, and frequencies of order
MHz, we obtain $\xi\lesssim \unit{1}{nm}$.\\
We can thus conclude that within the reach of current distance scales in trap setups,
it will be hard to observe the predicted effects of correlations, unless patch noise is the correct 
origin of anomalous heating.

\section{Experimental proposal} 
\begin{figure}[t!]
\includegraphics[width=\columnwidth]{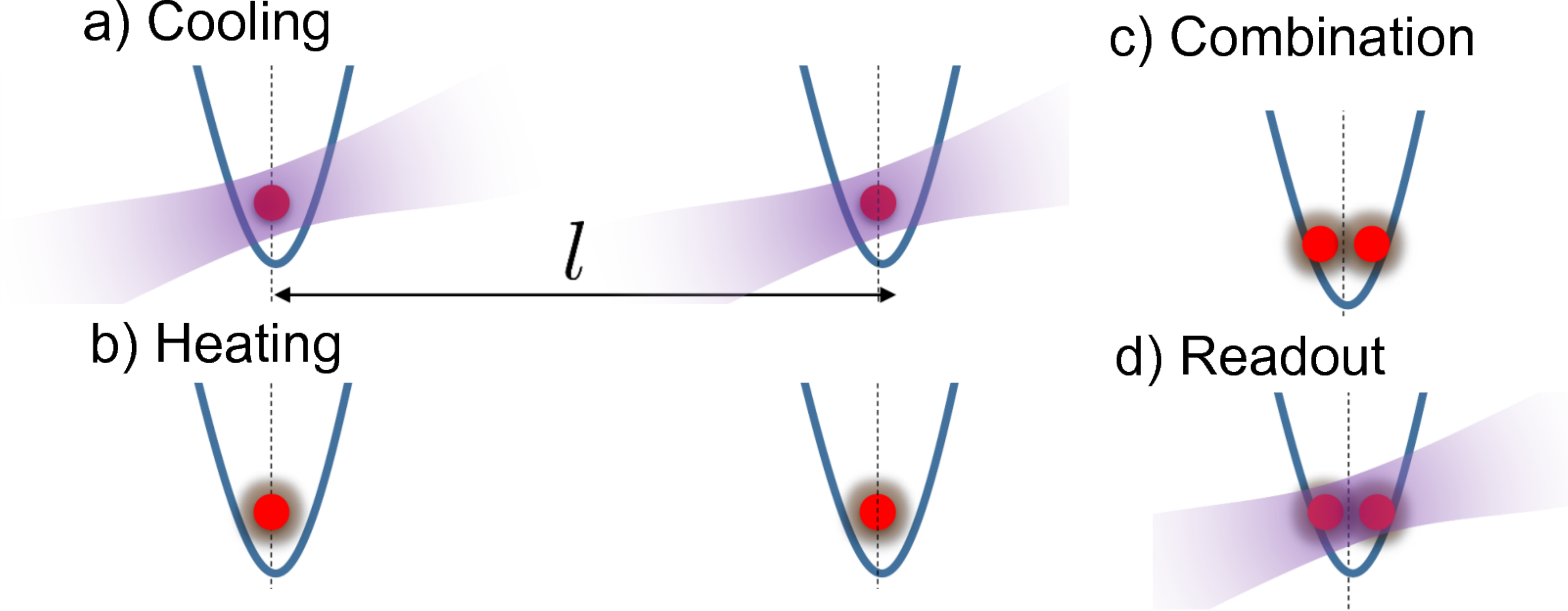}
\caption{{\it Experimental sequence.-}
The four radial normal modes of motion of a two-ion system are first cooled 
close to the ground state \cite{16Lechner}. After being exposed to interactions with the environment, the 
ions are combined in a common potential well. There the motional states of 
all four modes are read by coupling to the internal electronic states \cite{16Alonso},
giving access to the desired normal modes' heating rates $\Gamma_\pm$.}
\label{exp_prop}
\end{figure}

In what follows we present an experimental 
routine designed to measure the noise crossover from radiative sources on the 
surfaces of ion-trap electrodes (Fig. \ref{exp_prop}).
With two ions, a straightforward approach is to measure the heating rate 
$\Gamma_\pm$ of the different normal modes as a function of the distance $l$ 
between the ions, while keeping 
the ion-electrode separation $d$ constant. From equation (1) we obtain the noise 
spectral densities $S_{+}$ and $S_{-}$, and from them we calculate $S_{\rm 
self}=(S_{+}+S_{-})/2$ and $S_{\rm cross}=(S_{+}-S_{-})/2$.
As explained earlier, for the pure CB case the noise ratio $S_{\rm cross}/S_{\rm 
self}=1$, and only the COM mode will heat up; for pure aCB $S_{\rm 
cross}/S_{\rm self}=-1$, only the relative motion heats up; for SB 
$S_{\rm cross}/S_{\rm self}=0$, they will both get excited according to the 
spectral noise-density present at the modes' frequencies. 
Such an experiment can be carried out in a linear trap, where radial modes are 
orthogonal to the trap axis. In our scheme, the heating rates of the radial COM
and rocking modes are measured for different ion separations. The plots in Fig. 
\ref{fig3} show that an interesting range for the ratio $l/d$ goes from 0.5 to 
10 - throughout 
this range the noise ratio varies strongly in all the cases we have simulated. 
For an ion-trap height $d=\unit{50}{\micro m}$ we should therefore be able to 
vary $l$ from 25 to 
$\unit{500}{\micro m}$. With realistic experimental parameters 
the coupling rates between the motional modes of two ions spaced by 
$\unit{500}{\micro m}$ cannot be expected 
to exceed $\sim\unit{1}{Hz}$ \cite{14Krauth}, making it impossible to spectrally 
resolve the normal modes at mega-hertz frequencies. One way of overcoming this limitation is to let the ions heat up while separated 
and then bring them together to the same potential well to determine the 
normal-mode states, as in 
Fig. \ref{exp_prop}. Ion-chain splitting and recombination operations have been 
successfully carried out with negligible effects on the radial degrees of 
freedom \cite{16Kaufmann}.
When the ions crystallize in the same trap, the normal modes are easily 
resolvable \cite{98Wineland2}.

In order to resolve the relevant features in figure \ref{fig3} it suffices with an 
absolute uncertainty of the $S_{\rm cross}/S_{\rm self}$ ratio $\delta S_\text{ratio}\lesssim 0.1$. 
Assuming a relative uncertainty $\delta\Gamma_+/\Gamma_+=\delta\Gamma_-/\Gamma_- = \epsilon$ 
on the determination of the normal-mode heating-rates, as well as no correlations between 
$\delta\Gamma_+$ and $\delta\Gamma_-$, we find that 
$\delta S_\text{ratio} = \epsilon\sqrt{1+S_\text{ratio}^2}\frac{\sqrt{\Gamma_+^2+\Gamma_-^2}}{\Gamma_++\Gamma_-} \leq \epsilon\sqrt{2}$. 
With uncertainties $\epsilon \lesssim 5\%$, which are experimentally feasible, we expect 
the crossovers predicted in table \ref{tab:table1} to be clearly resolvable.\\
\indent As mentioned before, directionality effects have 
been observed after treating electrode 
surfaces with ion bombardment with a well-defined angle, and might be indicative 
of the microscopic origin of the noise sources \cite{PrivateHite}. Our scheme is 
sensitive to such 
anisotropy, which would lead to measurable differences in the spectral noise 
densities for the different radial axes.

\section{Conclusion and outlook} Microscopic models for anomalous heating are based on dipolar 
sources with different characteristics: frequency scalings have been a major 
concern to distinguish among them, but insufficient
attention has been paid to their geometric characteristics. Here we  find that the normal-mode heating-rates of two ions experience a previously unknown crossover that can be used to distinguish mean dipole
orientations and dipole-dipole correlations (or patch sizes). We provide estimates of the latter for 
well known models in the literature and show their effect for one- and two-ions configurations.
We also propose an experiment which is feasible with current state-of-the-art setups.
The idea of exploring spatial characteristics of the noise with more than one ion is left as 
a new tool for future investigations, and has interesting consequences. For example, in a coupled
chain of $N$ ions (appendix \ref{appendix:appH}) the most noise-resistant normal modes are odd ones,
and should be the ones used as quantum information buffers. Further, the availability of many normal modes
could potentially give finer details on geometric features of dipole arrangements and correlations.

\begin{acknowledgments}
JA thanks Dustin Hite for discussions. This work has been supported by the EU through the H2020 Project QuProCS (Grant 
Agreement 641277), by
MINECO/AEI/FEDER through projects NoMaQ FIS2014-60343-P, QuStruct 
FIS2015-66860-P and EPheQuCS FIS2016-78010-P. The research is partly based upon work supported by the Office of 
the Director of National Intelligence (ODNI), Intelligence Advanced Research Projects Activity (IARPA), via
the U.S. Army Research Office grant W911NF-16-1-0070. The views and conclusions contained herein are those of
the authors and should not be interpreted as necessarily representing the official policies or endorsements, 
either expressed or implied, of the ODNI, IARPA, or the U.S. Government. The U.S. Government is authorized 
to reproduce and distribute reprints for Governmental purposes notwithstanding any copyright annotation 
thereon. Any opinions, findings, and conclusions or recommendations expressed in this material are those 
of the author(s) and do not necessarily reflect the view of the U.S. Army Research Office.

\end{acknowledgments}

\appendix

\section{Lindblad equation for 2 and N ions}
\label{appendix:appA}

Expanding the interaction Hamiltonian $q\phi(\vec{R})$ as in Appendix B around 
$\vec{r}=\{0,d,0\}$  (with the ion quantum fluctuations around that position 
$\delta \vec{r}$)
yields $H=\vec{\delta r}\cdot\vec{\nabla}\phi(\vec{r})=-\delta \vec{r}\cdot\vec{E}(\vec{R})$. Let us concentrate for simplicity on the interaction
(and resultant noise) along $x$
and drop the $\delta$; thus the interaction energy for two ions is 
$H_I=-x_1E_x(\vec{r}_1)-x_2E_x(\vec{r}_2)$, and similarly for $N$ ions we have 
$\sum_{i=1}^N x_iE_x(\vec{r}_i)$.
We assume that in general the ions are coupled by direct Coulomb 
interaction, so they will form a set of $N$ normal modes $Q_i=\sum_j f_{i,j} 
x_j$ with eigenfrequencies $\Omega_i$. 
Any perturbative noise calculation that we do must be referred to that eigenset (see \cite{breuer}). Let us rewrite the interaction Hamiltonian
\begin{eqnarray}
H_I&=&-\sum_{i=1}^N x_iE_x(\vec{r}_i)=-\sum_{i=1}^N\sum_{j=1}^N (f^T)_{i,j}Q_j 
E_x(\vec{r}_i)\nonumber\\
&=&-\sum_{j=1}^N \tilde{E}_x^{(j)} Q_j,
\end{eqnarray}
with the new `electric noises' $\tilde{E}_x^{(j)}=\sum_{i=1}^N 
f_{j,i}E_x(\vec{r}_i)$, and where we have used the fact that the transformation matrix $f$ to 
normal modes is orthogonal (the inverse is its transpose).
In the interaction picture the system variables $Q_j$ rotate as 
\begin{equation}
\sqrt{\frac{\hbar}{2m\Omega_j}}(A_j e^{-i\Omega_j t}+A_j^\dagger e^{i\Omega_j 
t})
\end{equation}
and we can already calculate heating rates in two ways: either we use the usual 
argument of obtaining the probability to jump 
from $|0\rangle$ to $|1\rangle$ in the Fock
basis of a given eigenmode, {\it or} we obtain a master equation for the set of 
eigenmodes. The first one assumes ground state cooling, while the second is 
generic.\\

A {\it first approach} which (naively) applies single heating rates (see e.g. a derivation in appendix A of \cite{blatt}) to
each eigenmode yields $$\Gamma_{0\to 1}^{(j)}:=\Gamma_j=\frac{e^2}{4m\hbar 
\Omega_j}S_\text{E}(\Omega_j),$$ with 
$$S_\text{E}(\Omega_j)=2\int_{-\infty}^\infty d\tau e^{-i\Omega_j \tau}\langle 
\tilde{E}_x^{(j)}(\tau)\tilde{E}_x^{(j)}(0)\rangle.$$
The correlator can be expanded$$\langle 
\tilde{E}_x^{(j)}(\tau)\tilde{E}_x^{(j)}(0)\rangle=\sum_{k,l}f_{j,k}f_{j,l} 
\langle E_x(\vec{r}_k,\tau)E_x(\vec{r}_l)\rangle.$$

In the case of two ions the matrix $f$ is
$$f=\frac{1}{\sqrt{2}}\begin{pmatrix}
    1 & 1 \\
   1 & -1
  \end{pmatrix}$$
or simply, $\tilde{E}_x^{(\pm)}=[E_x(\vec{r}_1)\pm E_x(\vec{r}_2)]/\sqrt{2}$, 
with $+/-$ corresponding to center of mass/stretch modes, and also to $Q_{1/2}$, 
as intuition tells. Finally, the noise kernels
suffered by center of mass and stretch modes are
\begin{widetext}
\begin{equation}
\langle \tilde{E}_x^\pm(\tau)\tilde{E}_x^\pm(0)\rangle=\frac{1}{2}\left[\langle 
E_x(\vec{r}_1,\tau)E_x(\vec{r}_1,0)\rangle+\langle 
E_x(\vec{r}_2,\tau)E_x(\vec{r}_2,0)\rangle\pm\langle 
E_x(\vec{r}_1,\tau)E_x(\vec{r}_2,0)\rangle
\pm\langle E_x(\vec{r}_2,\tau)E_x(\vec{r}_1,0)\rangle\right],
\end{equation}
\end{widetext}
which can seen to consist of a {\it self-damping} part (first two terms) and a 
{\it cross-damping} part (last two terms).\\

A generically correct {\it second approach}, is to derive the full dissipator in the Lindblad equation 
\cite{breuer} 
\begin{equation}
D(\rho)=\sum_\omega\sum_{\alpha,\beta} 
\gamma_{\alpha,\beta}(\omega)\left(A_\beta \rho A_\alpha^\dagger-\frac{1}{2}\{ 
A_\alpha^\dagger A_\beta,\rho \}   \right)
\end{equation}
with $\omega$ spanning $\Omega_\pm$ and  
$A_\alpha=A_+,A_-,A_+^\dagger,A_-^\dagger$ are the COM and stretch modes' ladder 
operators. The basic difference with the previous approach is that, in addition to the (correctly predicted) heating rates 
$\Gamma_j$ written above, there appear now cooling rates too, {\it but also} 
cross-heating 
and cross-cooling rates which are neglected in the previous approach. The 
appearance of cooling is obvious from the time-symmetry of the evolution, but we 
need not care 
about it since we are interested in an experimental routine where we cool the 
normal modes at the beginning of each experimental run. 

Interestingly, and sometimes overlooked in the literature, there are cross-terms 
in the dissipator which couple the normal modes, with kernels of the type
\begin{widetext}
\begin{equation}
\langle\tilde{E}_x^+(\tau)\tilde{E}_x^-(0)\rangle=\frac{1}{2}\left[\langle 
E_x(\vec{r}_1,\tau)E_x(\vec{r}_1,0)\rangle-\langle 
E_x(\vec{r}_2,\tau)E_x(\vec{r}_2,0)\rangle+\langle 
E_x(\vec{r}_2,\tau)E_x(\vec{r}_1,0)\rangle-\langle 
E_x(\vec{r}_1,\tau)E_x(\vec{r}_2,0)\rangle\right].
\end{equation}
\end{widetext}
Typically these terms can be neglected because the normal modes have different frequencies and thus these terms rotate fast [$\Omega_+-\Omega_-\gg \gamma(\Omega_\pm)$].
However, when the coupling to the environment is strong enough, or the normal modes frequencies small enough (our case here
because ions are heated when they are far apart and feel almost no Coulomb coupling), these terms become
important [$\Omega_+-\Omega_-\sim \gamma(\Omega_\pm)$]. 

Luckily enough, for a sufficiently homogeneous sample the first two terms will cancel out 
(through similar noise conditions in the two ions positions) and also the last two terms. If this is fulfilled, the Lindblad 
dissipator separates into two independent dissipation channels,
one for each normal mode. Thus finally, the first approach seems to be 
sufficient if we consider ground state cooled normal modes, and that the 
electrode is more or less homogeneously (though random microscopically) populated by 
adsorbed atoms.\\

{\it Uncoupled ions.-} In the proposed experimental implementation we cool the 
motion of two ions separated by $l$, where they can be only very weakly coupled. 
If we consider the ion-ion coupling to be negligible [$\Omega_+-\Omega_-\ll \gamma(\Omega_\pm)$], the Lindblad equation \cite{breuer} is 
$$D(\rho)=\sum_\omega\sum_{\alpha,\beta} 
\gamma_{\alpha,\beta}(\omega)\left(a_\beta \rho a_\alpha^\dagger-\frac{1}{2}\{ 
a_\alpha^\dagger a_\beta,\rho \}   \right)$$
with $\alpha,\beta=a_1,a_2,a_1^\dagger,a_2^\dagger$ the usual 
creation-annihilation operators for the axial ions motion. This dissipator can 
be diagonalized in the basis $a_\pm=(a_1\pm a_2)/2$, if we have
$\gamma_{1,1}=\gamma_{2,2}\hat{=}\gamma_{\textrm{self}} $. It yields independent 
dissipation for the C.O.M. and stretch modes, as was expected by symmetry ({\it 
i.e. as before but the Lindbladian
does not couple center of mass and stretch modes}). Respectively they dissipate 
with
$\gamma_\pm=\gamma_{\textrm{self}}\pm\gamma_{1,2}$. These coefficients are again 
the Fourier transform of the time-correlation functions
\begin{equation}
\gamma_{\alpha,\beta}(\omega)=\mathcal{F}\left(\left< E_x(\vec{r}_\alpha,t) 
E_x(\vec{r}_\beta)\right>\right)
\end{equation}

{\it Summary.-} All this discussion was intended to show all the 
pitfalls that exist when considering the generic problem of a coupled two-body 
dissipative system: there is a regime where cross-coupling between normal modes exists.
This regime however is not of significance if we assume enough homogeneity of the 
noise sources, which leads to independent heating rates for the normal modes.

We thus arrive to an intuitive picture: whenever we have that `cross'$\sim\left< 
E_x(\vec{r}_1,t) E_x(\vec{r}_2,0)\right>$ is similar to 
`self'$\sim\left< E_x(\vec{r}_1,t) E_x(\vec{r}_1,0)\right>=\left< 
E_x(\vec{r}_2,t) E_x(\vec{r}_2,0)\right>$, we will have what is normally called 
a {\it common bath} or a spatially-correlated environment, and the 
stretch mode will not dissipate. Note also that the sign of the cross term is 
very important: if it is positive it will induce higher dissipation for the COM, 
while when negative the stretch will suffer more.

Hence, we call for short (in analogy with common-use nomenclature)
\begin{eqnarray}
S_{\rm{cross}}&=&\frac{1}{2}(\left< E_x(\vec{r}_1,t) E_x(\vec{r}_2,0)\right>+\left< E_x(\vec{r}_2,t) 
E_x(\vec{r}_1,0)\right>)\nonumber\\
S_{\rm{self}}&=&\frac{1}{2}(\left< E_x(\vec{r}_1,t) E_x(\vec{r}_1,0)\right>+\left< E_x(\vec{r}_2,t) 
E_x(\vec{r}_2,0)\right>)\nonumber
\end{eqnarray}
and compare their magnitudes and relative sign in the main text.

\section{Dipole geometric functions}
\label{appendix:appB}
The spatial functions describing the interaction of one dipole $\vec{\mu}$ with 
the ion motion in a given axis are given here. Noting that the ion is at 
$\vec{r}=\{x,y,z\}$ (fluctuating 
close to the point $\{0,d,0\}$), the dipole is at $\vec{r}_d=\{x_d,0,z_d\}$, the 
distance is defined as $\vec{R}= \vec{r}-\vec{r}_d$ and the electric potential 
between both is 
$$\phi=\frac{1}{4\pi\epsilon_0}\frac{\vec{\mu}\cdot\vec{R}}{|\vec{R}|^3}$$
we can easily obtain the total electric field in any direction 
$\vec{E}=-\vec{\nabla} \phi$. To obtain the noise felt by the ion in one of its 
eigenmotions, axial or radial, we need to calculate the corresponding
component of that electric field; we will also write down the expressions when 
assuming that the dipoles are pointing {\it only} along a given direction. We 
define the dipole functions
as $g_n(\vec{r})=-(4\pi\epsilon_0/|\vec{\mu}|) \partial_n\phi(\vec{r})$, after 
expanding the potential around $\vec{r}\simeq\{0,d,0\}$:

{\it Noise along x motion:}
\begin{eqnarray}
g_x(\vec{r})&=&\frac{d^2 - 2 x_d^2 + z_d^2}{(d^2 + x_d^2 + 
z_d^2)^{5/2}}\hspace{1cm},\hspace{1cm} \vec{\mu}=\mu \hat{u}_x\nonumber\\
g_x(\vec{r})&=&\frac{3 d x_d}{(d^2 + x_d^2 + 
z_d^2)^{5/2}}\hspace{1cm},\hspace{1cm} \vec{\mu}=\mu \hat{u}_y\nonumber\\
g_x(\vec{r})&=&-\frac{3 x_d z_d}{(d^2 + x_d^2 + 
z_d^2)^{5/2}}\hspace{1cm},\hspace{1cm} \vec{\mu}=\mu \hat{u}_z\nonumber
\end{eqnarray}

{\it Noise along y motion:}
\begin{eqnarray}
g_y(\vec{r})&=&\frac{3dx_d}{(d^2 + x_d^2 + 
z_d^2)^{5/2}}\hspace{1cm},\hspace{1cm} \vec{\mu}=\mu \hat{u}_x\nonumber\\
g_y(\vec{r})&=&-\frac{2d^2 - x_d^2 - z_d^2}{(d^2 + x_d^2 + 
z_d^2)^{5/2}}\hspace{1cm},\hspace{1cm} \vec{\mu}=\mu \hat{u}_y\nonumber\\
g_y(\vec{r})&=&\frac{3 d z_d}{(d^2 + x_d^2 + 
z_d^2)^{5/2}}\hspace{1cm},\hspace{1cm} \vec{\mu}=\mu \hat{u}_z\nonumber
\end{eqnarray}

{\it Noise along z motion:}
\begin{eqnarray}
g_z(\vec{r})&=&\frac{-3 x_dz_d}{(d^2 + x_d^2 + 
z_d^2)^{5/2}}\hspace{1cm},\hspace{1cm} \vec{\mu}=\mu \hat{u}_x\nonumber\\
g_z(\vec{r})&=&\frac{3 d z_d}{(d^2 + x_d^2 + 
z_d^2)^{5/2}}\hspace{1cm},\hspace{1cm} \vec{\mu}=\mu \hat{u}_y\nonumber\\
g_z(\vec{r})&=&\frac{d^2 +x_d^2 - 2z_d^2}{(d^2 + x_d^2 + 
z_d^2)^{5/2}}\hspace{1cm},\hspace{1cm} \vec{\mu}=\mu \hat{u}_z\nonumber
\end{eqnarray}

In the case of 2 ions, their positions will now be $\vec{r}_1=\{-l/2,d,0\}$ and $\vec{r}_2=\{l/2,d,0\}$ 
and we can use the former expressions by substituting $x_d\to x_d\pm l/2$ respectively.

\section{Origin of cross-noise vanishing}
\label{appendix:appC}
We have argued that the cross-noise vanishes for some ion motions and dipoles 
orientations. Let us take for example motion along $x$ and dipoles pointing 
normal to the electrode ($\vec{\mu}=\mu\hat{u}_y$). Considering
for the moment a collection of uncorrelated, homogeneously distributed dipoles, 
we have that the cross-noise of two ions sitting at $\{-l/2,d,0\}$ and 
$\{l/2,d,0\}$ is proportional to
\begin{figure}[h!]
\includegraphics[width=0.65\columnwidth]{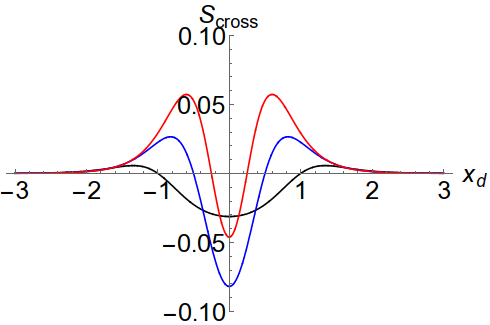}\\
\includegraphics[width=0.65\columnwidth]{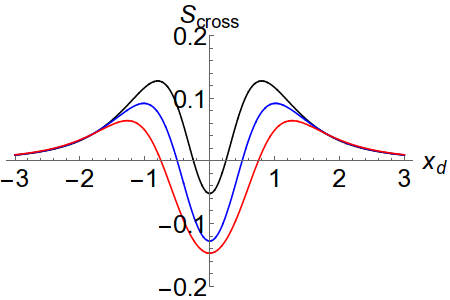}
\caption{(Top) Spatial dependence of cross-noise for $x$ motion and dipoles 
oriented along $y$. (Bottom) Same plot for monopolar sources. Depending on the 
ratio $l/d=2,1,0.5$ (black, blue, red) the shape enclose a negative, null or 
positive area, whence the noise crossover 
behavior.}
\label{d=l}
\end{figure}
$$\frac{(x_d-l/2)(x_d+l/2)}{(d^2+z_d^2+(x_d-l/2)^2)^{5/2}
(d^2+z_d^2+(x_d+l/2)^2)^{5/2}}.$$ Integrating this function along $z_d$ yields 
always a finite value, however the behavior along $x_d$ is like an M, 
as seen in figure~\ref{d=l} (top). The area enclosed by this curve can be 
positive, negative or zero depending on the ratio $d/l$, with the change of sign 
occurring near $d=l$.
For other directions of motion and dipole orientations, using the functions 
$g_n(\vec{r})$ given in Appendix B, it is easy to deduce the properties that we 
have summed up in figure~\ref{fig3} and Table I.

What about monopolar charges? In that case, the electric potential $\phi\propto 
1/|\vec{R}|$ is even (instead of odd) under reflection $\vec{R}\to -\vec{R}$. 
Still in this case, we have a crossover, see figure~\ref{d=l} (bottom).
Take as before $x$-motion and dipoles along $y$, the cross-noise is here 
proportional to
$$\frac{(x_d-l/2)(x_d+l/2)}{(d^2+z_d^2+(x_d-l/2)^2)^{3/2}
(d^2+z_d^2+(x_d+l/2)^2)^{3/2}},$$ which is the same as above only that instead 
of $5/2$ we have $3/2$ exponents. The difference in crossover is that for 
monopoles it occurs
at $l/d\simeq 2$.

\section{Pure anti-common bath regime}
\label{appendix:appD}
\begin{figure}[h!]
\includegraphics[width=0.7\columnwidth]{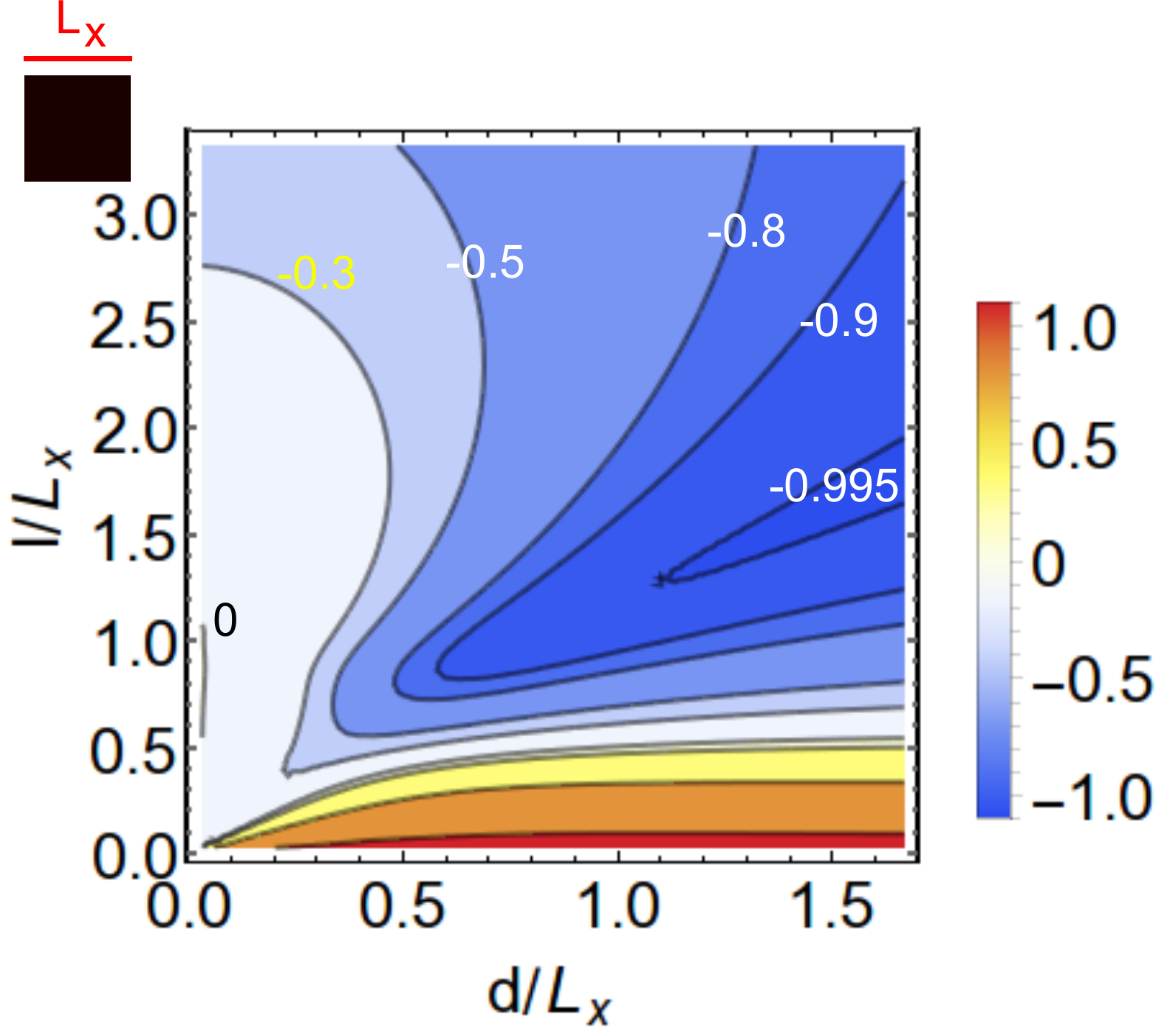}\\
\includegraphics[width=0.7\columnwidth]{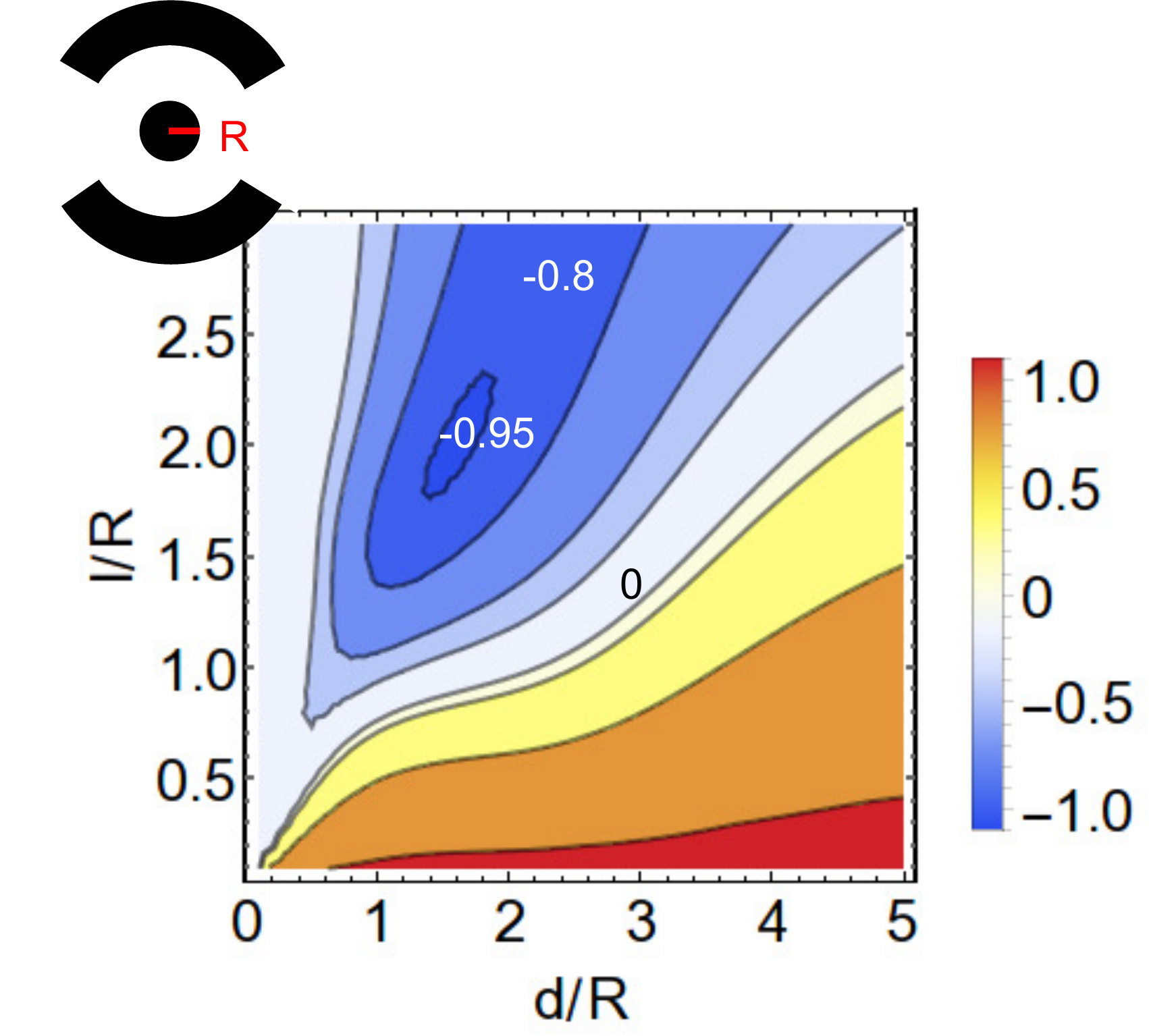}
\caption{Noise ratios $S_{\rm cross}/S_{\rm self}$ for the $x$-motion of 2 ions 
above two different geometries, with homogeneous distribution of dipoles 
pointing normal to the surface: (Top) square electrode $L_x=L_z$,
when the rest of $y=0$ plane is  dielectric (or empty), and (Bottom) a stylus 
trap as in \cite{wineland2017} with inner disk electrode of radius $R$ and an 
outer ring from $3R$ to $5R$; 
further apart there are 4 disks, but for the distances considered here they do 
not affect the ions; the spaces between electrodes act as empty spaces.}
\label{aCB}
\end{figure}

It is intuitive that when two coupled units are close to each other and far away 
from a noisy environment they will experience a common bath (CB), and this is 
what has been predicted in the main text. But, what about
the opposite, is there a regime where the stretch mode is the only dissipative 
degree of freedom? In the main text figures we have seen that the minimum 
$S_{\rm cross}/S_{\rm self}$ was around $-0.4$.
We investigate this in figure \ref{aCB} for two geometries: a square finite 
electrode, and a stylus trap as in \cite{wineland2017}. For the square electrode 
a value of almost -1 is reached around 
$\{d,l\}=\{1.2L_x,1.4L_x\}$. A value lower than $-0.9$ is reached 
$\{d,l\}\simeq\{L_x,L_x\}$. For the stylus trap a lowest value of $\simeq -0.95$ 
is reached at reasonable distances $\{d,l\}=\{1.5R,2R\}$, 
although the absence of a common RF null for two ions in this configuration 
might require active stabilization of the RF-drive amplitude, whose noise could 
otherwise mask anomalous heating.

\section{1-ion noise level for different dipole orientations}
\label{appendix:appE}
Dipole-dipole interactions, as that caused by a surface dipole and the displaced charged ion from its
equilibrium position, have a preferred direction. It is hence to be expected that different orientations
of dipole sources will yield different noise levels even for a 1-ion configuration. Next we plot the simplest
situation in which a uniform planar infinite electrode is filled with dipoles oriented along the 3 
possible directions: an ion's motion along $x$ feels highest noise levels when dipoles are along $\hat{u}_y$, while
for in-plane dipole orientations,  $\hat{u}_x$ produces a noise 4 times higher than $\hat{u}_z$.
The asymmetry between the 3 directions is a direct consequence of the dipole dipole interaction form
$3(\vec{m}_1\cdot\hat{r})(\vec{m}_2\cdot\hat{r})-\vec{m}_1\cdot\vec{m}_2$, with $\vec{m}_1$ the ion displacement [along $\hat{u}_x$], $\hat{r}$ the 
ion-dipole unit vector [along $\hat{u}_y$] and $\vec{m}_2$ the dipole moments on the surface.
\begin{figure}[h!]
\includegraphics[width=0.7\columnwidth]{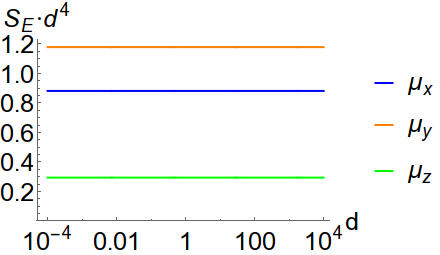}
\caption{Noise felt by a single 
ion in its motion along $x$ when dipoles are pointing in different directions, with
an electrode of infinite size arranged as in figure~\ref{fig1}.
The noise has been normalized by the usual scaling $d^{-4}$ for clarity. The 
electrode is of infinite size. Noise along the 
trap axis is clearly dominant.}
\label{dipole1ion}
\end{figure}
In a similar spirit, Schindler and coworkers \cite{Harmut2} showed (although assuming dipoles perpendicular to the electrode)
that one can use the asymmetry in noise levels to distinguish between technical noise and anomalous heating sources. Also, the idea 
of 1-ion noise sensing for studying stray fields is used in \cite{FSK}.

\section{Mean field dipole}
\label{appendix:appF}
The blurring out into a mean-field dipole in the presence of spatial 
dipole-dipole correlations can be understood as follows. We can rewrite the 
noise spectral density seen by 1 ion as
the Fourier transform of
\begin{eqnarray}
S&=&\langle E_x(\tau)E_x(0)\rangle\nonumber\\
&=&\sum_{i,j}\left(\frac{1}{4\pi\epsilon_0}\right)^2 \left< 
\mu_i(t)\mu_k(0)\right>g_x(\vec{r}_i)g_x(\vec{r}_j)
\end{eqnarray}
where $\vec{r}_i$ is the distance between dipole $i$ and the ion. As explained 
in the main text, the spatial-temporal separability of 
$\left< \mu_i(t)\mu_k(0)\right>$ allows us to approximate it by $\left< 
\mu_i(t)\mu_i(0)\right>f(\vec{r}_i,\vec{r}_j)\simeq 
S_\mu(t)f(\vec{r}_i,\vec{r}_j)$
where we have assumed that temporal dipole fluctuations are similar for 
different spatial regions of the electrode, something rather reasonable. 
Hence, we can write
\begin{eqnarray}
S&=&\langle E_x(\tau)E_x(0)\rangle\nonumber\\
&\simeq& \left(\frac{1}{4\pi\epsilon_0}\right)^2 
S_\mu(t)\sum_{i,j} g_x(\vec{r}_i)g_x(\vec{r}_j)f(\vec{r}_i,\vec{r}_j)
\end{eqnarray}
Further, we can assume that the spatial dipole-dipole correlation profile 
$f(\cdot)$ is sufficiently translational invariant in the regimes of interest 
(e.g. ions do not approach too close
the borders of electrodes) and thus only depends on the absolute distance 
between dipoles: $f(\vec{r}_i,\vec{r}_j)\simeq f(|\vec{r}_i-\vec{r}_j|)$.
We can sum up the situation by writing the spatial dependence of the spectral 
noise as
$$\sum_{i,j}g_x(\vec{r}_i)g_x(\vec{r}_j)f(|\vec{r}_i-\vec{r}_j|).$$

If the function $f$ decays significantly for distances greater than $\xi$ as 
e.g. if $f(\vec{r}_i,\vec{r}_j)=e^{-|\vec{r}_i-\vec{r}_j|/\xi}$, we can define an 
effective, or `mean field', function 
$\tilde{g}_x(\vec{r}_i;\xi):=\sum_j g_x(\vec{r}_j)f(|\vec{r}_i-\vec{r}_j|)$ 
centered at $\vec{r}_i$ and averaged over a size $\xi$. The noise sum then 
becomes 
$\sum_i g_x(\vec{r}_i)\tilde{g}_x(\vec{r}_i;\xi)$, to be compared with the case 
of uncorrelated dipoles $\sum_i g_x(\vec{r}_i)^2$. 
Now recall that the $g$ functions are meaningful only in an area of order 
$\mathcal{A}\sim \mathcal{O}(d)$ around the ion position, so contributions
$g_x$ of dipoles far from this spot can be neglected. This allows us to compare
the region $\mathcal{A}_d$ of ion-dipoles influence, to the region size $\mathcal{A}_\xi$
defining the averages for the mean-field dipoles. We thus have the following situations:
\begin{itemize}
 \item $\xi\ll d$ ($\mathcal{A}_d\gg\mathcal{A}_\xi$): Mean-field dipole is seen as an effectively individual 
dipole and so $\tilde{g}_x(\vec{r}_i;\xi)\to g_x(\vec{r}_i)$. This is the 
uncorrelated dipoles case, which is known to scale as $d^{-4}$.
 \item $\xi\geq d$ : When $\xi$ increases, we reach a point where the averaging region 
 $\mathcal{A}_\xi$ becomes bigger than $\mathcal{A}_d$, but every dipole lying out of it
 has a negligible influence on the ion. This means that making $\xi$ bigger yet will not modify the value
 of the mean-dipole function, which at this $\xi$ {\it saturates}.
\end{itemize}
 
 This argument explains why the scaling drops from $d^{-4}$ when $\xi\sim d$ is reached, however it does not explain
 the new scaling $d^{-1}$ for $d<\xi$. We can, however give an argument why the exponent is less than $4$:\\
 When $\mathcal{A}_\xi\geq\mathcal{A}_d$ the sum for the mean-field dipole includes the full region 
of influence see by the ion. If we do the brutal simplification $f(|\vec{r}_i-\vec{r}_j|)\simeq1$ [$\forall i,j\in \mathcal{A}_d]$, 
 $\sum_{i,j}g_x(\vec{r}_i)g_x(\vec{r}_j)f(|\vec{r}_i-\vec{r}_j|)\to \sum_{i\in 
\mathcal{A}} g_x(\vec{r}_i)\sum_{j\in \mathcal{A}}g_x(\vec{r}_j)=[\sum_{i\in 
\mathcal{A}} g_x(\vec{r}_i)]^2$ this sum, for dipoles pointing normal to 
 the surface, would yield simply $0$ (a constant, i.e. $d^0$).
 However if instead of taking $f(|\vec{r}_i-\vec{r}_j|)\to 1$, we use a linear expansion of the exponential 
function $f\sim 1-|\vec{r}_i-\vec{r}_j|/\xi$ the sum yields a scaling $d^{-1}$ 
as is  observed in the next Appendix and is consistent with correlated patch models \cite{guidoni}. It is not obvious that
in general the exact resulting exponent should be $-1$, but it seems intuitive that the strength of the scaling is tamed.

Further, for the specific case of next appendix [ion motion along $x$ and dipoles pointing along $y$] we can approximate the 
dipoles integration as follows: let us take the dipole geometric function from appendix B, which reads 
$$g_x(\vec{r})=\frac{3d x}{(d^ 2+x^ 2+z^ 2)^ {5/2}}$$
and approximate its biggest contribution which occurs for $\{x,d\}\ll d$ [this approximation usually gives a good estimate of
scalings with respect to $d$],
$$g_x(\vec{r})\sim 3x/d^ 4.$$
From now on we will drop numerical factors since we are interested only in the scalings. The noise felt by 1 ion is then
$$\sum_{i,j}g_x(\vec{r}_i)g_x(\vec{r}_j)f(\vec{r}_i,\vec{r}_j)\sim d^{-8}\sum_{i,j} x_i x_j f(\vec{r}_i,\vec{r}_j).$$ 
Transforming into integrals we have $$d^{-8}\int dx\int dx'\int dz\int dz' x x'f(\vec{r}_i,\vec{r}_j).$$
If dipoles are uncorrelated, $f(\vec{r}_i,\vec{r}_j)=\delta(\vec{r}_i,\vec{r}_j)$, and we have $d^{-8}\int dx\int dz x^2$, which gives
(recall that the important contribution from dipoles comes from an area of size $\sim d$) $d^{-4}$ as expected.

For the case of very big correlation length $\xi$, we can expand the exponential
\begin{widetext}
$$d^{-8}\int dx\int dx'\int dz\int dz' x x' \exp(\sqrt{(x-x')^2+(z-z')^2}/\xi)\sim-d^{-8}\int dx dx' dz dz' x x' (1-\sqrt{(x-x')^2+(z-z')^2}/\xi)$$
\end{widetext}
The part with the 1 gives 0 by symmetry, and the rest can be integrated by performing the change of variables $Q_\pm=(x\pm x')/\sqrt{2}$
and $Z_\pm=(z\pm z')/\sqrt{2}$, so
$$\sim d^{-8}\int dQ_- dQ_+ dZ_- dZ_+ (Q_+^2-Q_-^2)\sqrt{Q_-^2+Z_-^2}.$$
A final change of variables for the area enclosed by $Z_-$ and $Q_-$ (of order $d$) to polar $Q_-=r\cos\phi$, $Z_-=r\sin\phi$, gives
$$\sim d^{-8}\int dQ_+ dZ_+\int dr d\phi\ r (Q_+^2-r^2\cos\phi) r$$ which after simple integration yields $d^{-1}$. This highlights the power of approximating
the dipole functions in this way.


\section{Modified noise due to correlated dipoles}
\label{appendix:appG}
\begin{figure}[h!]
\includegraphics[width=0.8\columnwidth]{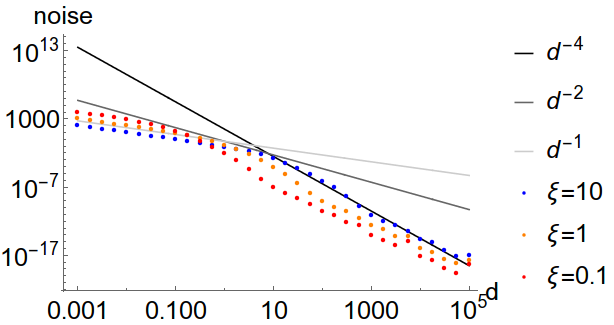}
\caption{Noise felt by 1 ion oscillating along $x$ when dipoles in the surface 
are pointing along $y$, with spatial dipole-dipole correlation distance $\xi$ .
The typical scaling $d^{-4}$ is broken for ion-electrode distances $d<\xi$, 
where it becomes $d^{-1}$. Distance is given in arbitrary units.}
\label{1ion}
\end{figure}

Here we study what is the effect of a finite extent $\xi$ of spatial 
dipole-dipole correlations on the characteristics of the noise crossover. For 
simplicity we take a profile function 
$$f(\vec{r}_i,\vec{r}_j)=e^{-|\vec{r}_i-\vec{r}_j|/\xi}$$
although for phononic-induced dipole vibrations of adatoms we should use the 
more realistic sinc($|\vec{r}_i-\vec{r}_j|/\xi)$, and for diffusion of adatoms 
the Kelvin function ${\rm Ker}_0(|\vec{r}_i-\vec{r}_j|/\xi)$.
In the case of {\it one ion} the noise follows, as is well known in the 
literature \cite{blatt}, a $d^{-4}$ behavior. However, for correlated dipoles, 
this scaling is modified for $d<\xi$, becoming $d^{-1}$ as can be seen in
figure \ref{1ion}.

\begin{figure*}[t!]
 \includegraphics[width=\textwidth]{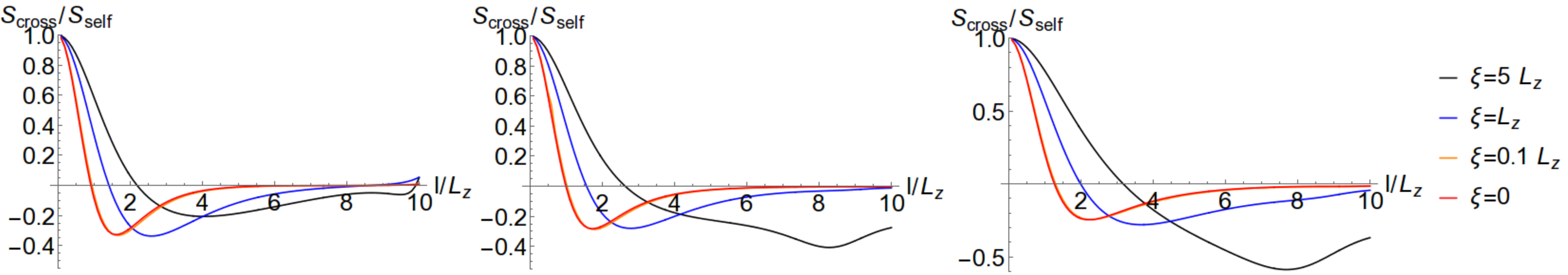}\\
 \includegraphics[width=\textwidth]{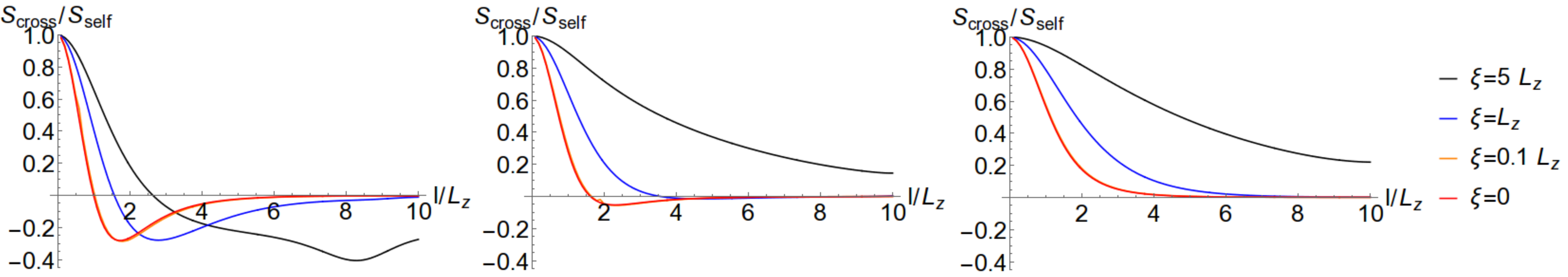}\\
 \includegraphics[width=\textwidth]{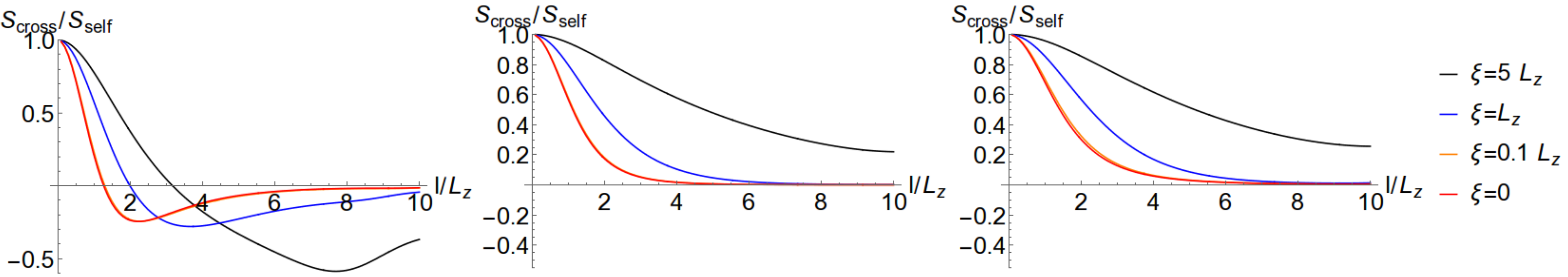}
\caption{Noise ratios $S_{\rm cross}/S_{\rm self}$ for 2 ions oscillating along 
$x$ (top), $y$ (middle) and $z$ (bottom),  when dipoles in the surface are pointing along $x,y,z$ (left, middle, 
right). Influence of spatial correlations is noticeable for $\xi>0.1L_z$.}
\label{xmotion}
\end{figure*}

We plot in figure \ref{xmotion} the crossover equivalents of figure~\ref{fig3} with correlated 
dipoles, taking the electrode configuration described in its caption.

\section{Chains of coupled ions}
\label{appendix:appH}
\begin{figure}[h!]
\includegraphics[width=\columnwidth]{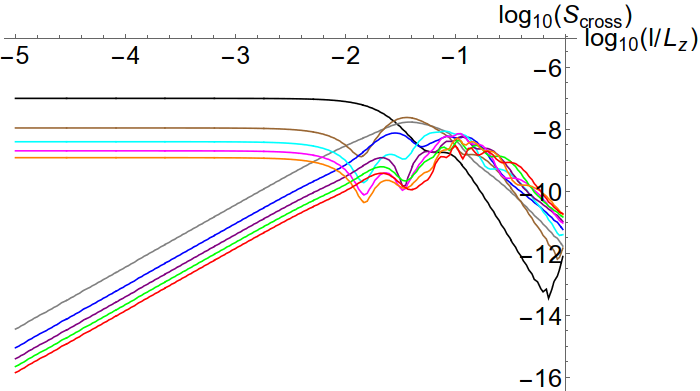}
\caption{Cross-noise (arbitrary units) along $x$-motion for a chain of 10 
strongly coupled ions when dipoles in the surface are pointing along $x$, as a 
function of the inter-ion distance $l\in[10^{-5}L_z,Lz]$.
The electrode has dimensions $L_x=10L_z$ and the ions are above the electrode at 
$d=L_z$. Black line is the mode with longest wavelength.}
\label{10ions}
\end{figure}
We briefly investigate what happens in a configuration where 10 ions are forming 
a Coulomb crystal. When they are strongly coupled we expect that normal modes of 
the chain can divide into sinusoidal waves with 2 kinds of parities: even 
or odd (under reflection in the center of the chain). It is intuitively expected 
that even modes will coupled strongly to a CB type of bath, while odd ones will do 
so for aCB types of baths. This is precisely what we observe in figure
\ref{10ions}: when ions get closer and closer to each other [$d=L_z$, $l\ll 
L_z$) they begin to see a pure CB configuration, and the noise suffered by odd 
modes vanishes while for even modes it stays high.
Further notice that when $l\simeq L_z$, i.e. when the ion chain almost occupies 
the full length $L_x$ of the electrode, the mode with longest wavelength (black 
line) sees the least noise. This is caused by ions being at the point 
nearest to the aCB regime, and thus the even symmetry of this mode makes it 
most isolated to noise. These features are of importance for the use of 
normal modes as buses for quantum information and for quantum simulations
where they provide effective spin-spin interactions among different ions 
\cite{porras}.


\begin{thebibliography}{10}
\bibitem{footnote} Even in cases where it is not spatio-temporal separable (e.g. adatom diffusion), its frequency spectrum is only important at the ion's 
frequency, which immediately singles out a purely spatial dependence.
\bibitem{note1} Normal modes orientation is important too, but it can be read out from the ion's coupling strength to a laser beam with a well-defined direction \cite{98Wineland2}.
\end{thebibliography}
\end{document}